\documentclass[12pt,oneside, a4paper]{article}

\ifx\pdfoutput\undefined
\usepackage[dvips,bookmarks=false]{hyperref}	
\else
\usepackage{hyperref}	
\fi
\hypersetup{colorlinks,bookmarksopen,bookmarksnumbered,citecolor=blue,
linkcolor=black,pdfstartview=FitH,urlcolor=blue}

\usepackage{graphicx}
\usepackage{amssymb}
\usepackage{cite}
\usepackage{bm}
\usepackage{indentfirst}
\usepackage{amsmath}
\usepackage{hhline}
\usepackage{multirow}

\allowdisplaybreaks

\newcommand{\1}{\mbox{1}\hspace{-0.25em}\mbox{l}}

\usepackage{ulem}

\begin{document}

\begin{titlepage}

\begin{flushright}
IPMU20-0131
\end{flushright}

\begin{center}

\vspace{1cm}
{\large\textbf{
Electric dipole moments in the extended scotogenic models
}%
 }
\vspace{1cm}

\renewcommand{\thefootnote}{\fnsymbol{footnote}}
Motoko Fujiwara$^{1}$\footnote[1]{motoko@eken.phys.nagoya-u.ac.jp}
,
Junji Hisano$^{1,2,3}$\footnote[2]{hisano@eken.phys.nagoya-u.ac.jp}
,
Chihiro Kanai$^{1}$\footnote[3]{ckanai@eken.phys.nagoya-u.ac.jp}
,
Takashi Toma$^{4,5}$\footnote[4]{toma@staff.kanazawa-u.ac.jp}

\vspace{5mm}

\textit{
$^1${Department of Physics, Nagoya University,\\ Furo-cho Chikusa-ku, Nagoya, 464-8602 Japan}\\
 $^2${Kobayashi-Maskawa Institute for the Origin of Particles and the Universe,\\ Nagoya University,Furo-cho Chikusa-ku, Nagoya, 464-8602 Japan}\\
 $^3${Kavli IPMU (WPI), UTIAS, University of Tokyo, Kashiwa, 277-8584, Japan}\\
 $^4${Institute of Liberal Arts and Science, Kanazawa University,\\ Kakuma-machi, Kanazawa, 920-1192 Japan}\\
 $^5${Institute for Theoretical Physics, Kanazawa University, Kanazawa, 920-1192 Japan}\\
}

\vspace{8mm}

\abstract{
Electric dipole moments (EDMs) of charged leptons arise from a new source of CP violation in the lepton sector.
In this paper, we calculate the EDMs of the charged leptons in the minimal scotogenic model with two singlet fermions, and 
the models extended with one or two triplet fermions instead of the singlet fermions, 
taking into account the constraints of the neutrino oscillation data, the charged lepton flavor violation and perturbative unitarity bound for the Yukawa couplings. 
We show that the hybrid model with one singlet and one triplet fermions predicts an electron EDM larger than the other models in both normal and inverted neutrino mass hierarchy. 
We find some parameter space has already been ruled out by the current upper bound of the electron EDM and further parameter space can be explored by future experiments. 
}

\end{center}
\end{titlepage}

\renewcommand{\thefootnote}{\arabic{footnote}}
\newcommand{\bhline}[1]{\noalign{\hrule height #1}}
\newcommand{\bvline}[1]{\vrule width #1}

\setcounter{footnote}{0}

\setcounter{page}{1}

\section{Introduction}
\label{sec:1}
It has been confirmed through the neutrino oscillation experiments that neutrinos have non-zero masses and large mixing angles compared to the quark sector. 
A common and simple framework generating small neutrino masses is the seesaw mechanism~\cite{Minkowski:1977sc, Yanagida:1979as, GellMann:1980vs}.
There are three types of the seesaw mechanisms, which are called type-I, type-II, and type-III seesaw models. 
Some $SU(2)_L$ singlet or triplet fermions are introduced to the Standard Model (SM) for the type-I and type-III seesaw models, 
while an $SU(2)_L$ triplet scalar with non-zero hypercharge is introduced for the type-II seesaw models.

On the other hand, the existence of dark matter in the universe is apparent from the astrophysical observations such as the rotation curves of spiral galaxies, 
cosmic microwave background, structure formation of the universe, and the collision of the Bullet Cluster. 
One of the simplest possibilities to accommodate these two different ingredients (namely, small neutrino masses and dark matter) is the radiative neutrino mass models. 
In this framework, neutrino masses at the tree level are forbidden by a symmetry which also stabilizes a dark matter candidate categorized as so-called Weakly Interacting Massive Particles (WIMPs). 
The typical mass range of WIMPs is $\mathcal{O}(100)~\mathrm{GeV}$ to $\mathcal{O}(10)~\mathrm{TeV}$, 
and thus it is testable by various experiments and observations such as particle colliders, underground experiments, and cosmic-rays.
Many radiative neutrino mass models have been proposed so far~\cite{Cai:2017jrq}, and in particular, the scotogenic model in which an inert scalar doublet and two singlet fermions are introduced to the SM is 
a minimal model~\cite{Ma:2006km}.
It is straightforward to extend the minimal scotogenic model with one or two triplet fermions instead of the singlet fermions 
similar to the relationship between the type-I and type-III seesaw models at the tree level. 
The model extended with two $SU(2)_L$ triplet fermions has rich phenomenology compared to the minimal model because of the existence of the extra charged fermions~\cite{Ma:2008cu, Chao:2012sz, vonderPahlen:2016cbw, Biswas:2018ybc}. 
One may feel that the hybrid model with a singlet fermion and a triplet fermion is less attractive since the model is rather ad-hoc.
However, this model can improve the gauge coupling unification in non-supersymmetric grand unified theories~\cite{Ma:2005he, Bajc:2006ia, Dorsner:2006fx}, which motivates us to consider the hybrid model as well~\cite{Hirsch:2013ola, Merle:2016scw, Rocha-Moran:2016enp, Choubey:2017yyn, Restrepo:2019ilz, Avila:2019hhv}. 

Precise measurements of electric dipole moments (EDMs) are rapidly being updated. 
The current upper bound of the electron EDM $|d_e|/e\leq1.1\times10^{-29}~\mathrm{cm}$ at $90\%$ confidence level has been reported by the ACME Collaboration~\cite{Andreev:2018ayy}, 
and the future sensitivity is expected to reach up to $|d_e|/e=\mathcal{O}(10^{-30})~\mathrm{cm}$~\cite{Kara:2012ay, edm_future}.
This bound and future sensitivity can test the framework of the radiative neutrino mass models. 

In this paper, we calculate the charged lepton EDMs in the minimal scotogenic model with two singlet fermions and 
its extensions where one or two triplet fermions with zero hypercharge are added instead of the singlet fermions. 
The EDMs in the minimal scotogenic model have already been calculated in Ref.~\cite{Abada:2018zra}. 
Here we show that in its extensions new diagrams contributing to EDMs appear, and we emphasize these give a dominant contribution. 
We write down all the Feynman diagrams relevant to the EDMs and give the complete analytic formulae of the charged lepton EDMs at two-loop levels in those models. 
It will turn out that some diagrams eventually do not give a contribution to the EDMs. 
In addition, we numerically evaluate the electron EDM predicted in these models, and investigate the possible maximum value taking into account some relevant constraints such as the neutrino oscillation experiments, the charged lepton flavor violating (LFV) processes such as $\mu\to e\gamma$ and the perturbative unitarity bound. 

The rest of the paper is organized as follows. 
In Section~\ref{sec:2}, we briefly review the minimal scotogenic model and its extensions. 
The radiatively induced neutrino masses are also calculated, and the neutrino Yukawa coupling is written down by the Casas-Ibarra parametrization. 
The experimental values of the neutrino mass eigenvalues and the mixing angles obtained by the global fit analysis are also briefly summarized. 
In Section~\ref{sec:3}, the branching ratios of the relevant charged LFV processes are calculated for each model focusing on the processes $\ell_\alpha\to\ell_\beta\gamma$. 
Section~\ref{sec:4} is devoted to calculate the charged lepton EDMs at the two-loop level in analytic ways.
Based on the analytic results, some numerical evaluation of the electron EDM and comparison with the current bound and future sensitivity are given in Section~\ref{sec:5}.
We give our conclusion in Section~\ref{sec:6}.

\section{The models}
\label{sec:2}
\subsection{The minimal scotogenic model and its extensions}
In the minimal scotogenic model, an inert doublet scalar $\eta$ and two singlet fermions $N$ are introduced to the SM. 
The $\mathbb{Z}_2$ symmetry is imposed to forbid neutrino mass terms at the tree-level. 
This symmetry also stabilizes the lightest $\mathbb{Z}_2$ odd particle, thus this particle can be a dark matter candidate. 
Then, small neutrino masses are radiatively induced by these new particles at the one-loop level.
In addition to this model, we consider the extensions of the model where one or two triplet fermions $\Sigma$ are introduced instead of the singlet fermions. 
The particle contents of the models we consider are summarized in Tab.~\ref{tab:1}. 
We name the models as in Tab.~\ref{tab:2}, namely the SS model for two singlet fermions, the ST model for one singlet and one triplet fermions, and the TT model for two triplet fermions. 

The kinetic term and Yukawa interactions of the new particles are given by
\begin{align}
\mathcal{L}=\frac{1}{2}\overline{\Psi_i}\left(iD\hspace{-0.26cm}/-m_i\right)\Psi_i
+\left(D_{\mu}\eta\right)^{\dag}\left(D^{\mu}\eta\right)-y_{i\alpha}\eta\overline{\Psi_i}P_LL_{\alpha}+\mathrm{H.c.},
\end{align}
where $\Psi_i=N_i$ or $\Sigma_i$ ($i=1,2$) and $\alpha=e,\mu,\tau$. 
Note that we define $\Psi_1\equiv N_1$ and $\Psi_2\equiv \Sigma_1$ for the ST model.
The scalar potential is given by
\begin{align}
\mathcal{V}=\mu_\Phi^2|\Phi|^2+\mu_{\eta}^2|\eta|^2+\frac{\lambda_1}{2}|\Phi|^4+\frac{\lambda_2}{2}|\eta|^4
+\lambda_3|\Phi|^2|\eta|^2+\lambda_4|\Phi^{\dag}\eta|^2
+\frac{\lambda_5}{2}\left[\left(\Phi^{\dag}\eta\right)^2+\left(\eta^{\dag}\Phi\right)^2\right],
\end{align}
where $\Phi$ is the SM Higgs doublet. 
All the parameters in the scalar potential are real without loss of generality.
After the electroweak symmetry breaking, the SM Higgs doublet $\Phi$ gets a vacuum expectation value (VEV) and is written as $\Phi=\left(0,\langle\Phi\rangle+h/\sqrt{2}\right)$ while 
the inert doublet scalar $\eta=(\eta^+,\eta^0)$ where $\eta^0=(H+iA)/\sqrt{2}$ is assumed not to have a VEV $\langle\eta\rangle=0$. 
Using the stationary conditions, the Higgs boson mass, the charged inert scalar mass, CP-even and CP-odd neutral scalar masses are written as
\begin{align}
m_h^2&=2\lambda_1\langle\Phi\rangle^2,\\
m_{\eta^+}^2&=\mu_\eta^2+\lambda_3\langle\Phi\rangle^2,\\
m_H^2&=\mu_\eta^2+\left(\lambda_3+\lambda_4+\lambda_5\right)\langle\Phi\rangle^2,\\
m_A^2&=\mu_\eta^2+\left(\lambda_3+\lambda_4-\lambda_5\right)\langle\Phi\rangle^2.
\end{align}
Note that the mass splitting between the CP-even and CP-odd neutral scalars is given by $m_H^2-m_A^2=2\lambda_5\langle\Phi\rangle^2$, which implies that the magnitude of the mass splitting is controlled by the coupling $\lambda_5$.
In the following, we assume the charged scalar $\eta^+$, the CP-even scalar $H$, and CP-odd scalar $A$ are degenerate 
in order to safely evade the constraint of the electroweak precision tests (the oblique parameters). 
This implies that the scalar couplings $\lambda_4$ and $\lambda_5$ are sufficiently small. 

 \begin{table}[t]
\begin{center}
\caption{New particle contents.}
\label{tab:1}
  \begin{tabular}{|c||c|c|c|}\hline
   & $\eta$ & $N$ & $\Sigma$\\\hhline{|=#=|=|=|}
   $SU(2)_L$      & $\bm{2}$ & $\bm{1}$ & $\bm{3}$\\\hline
   $U(1)_Y$       & $1/2$    & $0$      & $0$\\\hline
   $\mathbb{Z}_2$ & $-1$     & $-1$     & $-1$\\\hline
   Spin           & $0$      & $1/2$    & $1/2$\\\hline
  \end{tabular}
\end{center}
 \end{table}

\begin{table}
 \begin{center}
\caption{Number of fermions in each model.}
\label{tab:2}
  \begin{tabular}{|c|c|c|c|}\hline
 & SS model & ST model & TT model\\\hline
 Number of fermions & \multirow{2}{*}{($2$, $0$)} & \multirow{2}{*}{($1$, $1$)} & \multirow{2}{*}{($0$, $2$)}\\
($n_N$, $n_\Sigma$) & & &\\\hline
\end{tabular}
 \end{center}
\end{table}

Before moving on further, we count the number of physical CP-violating phases in our models. In the basis where $\Psi_i$ and the charged leptons are mass eigenstates and $\lambda_5$ is real, the $2\times 3$ Yukawa coupling matrix $y_{i\alpha}$ is generically  complex. Three phases in $y_{i\alpha}$ are unphysical since they can be removed by the field redefinition of $L_\alpha$. Thus, the three remaining phases are physical. 

\subsection{Neutrino masses and mixing angles}
In the models we consider here, the neutrino masses are generated at the one-loop level as same as the minimal scotogenic model~\cite{Ma:2006km}, and the mass matrix is given by
\begin{equation}
\left(m_{\nu}\right)_{\alpha\beta}=\sum_{i=1}^{2}\frac{y_{i\alpha}y_{i\beta}m_i}{2(4\pi)^2}
\left[
\frac{m_H^2}{m_H^2-m_i^2}\log\left(\frac{m_H^2}{m_i^2}\right)
-\frac{m_A^2}{m_A^2-m_i^2}\log\left(\frac{m_A^2}{m_i^2}\right)
\right].
\end{equation} 
In particular, when the mass splitting between $H$ and $A$ is small ($\lambda_5$ is small), 
the above mass formula is simplified as
\begin{equation}
\left(m_{\nu}\right)_{\alpha\beta}\approx
\sum_{i=1}^{2}\frac{y_{i\alpha}y_{i\beta}m_i}{(4\pi)^2}
\frac{\lambda_5\langle \Phi\rangle^2}{m_{\eta^0}^2-m_i^2}
\left[
1-\frac{m_i^2}{m_{\eta^0}^2-m_i^2}\log\left(\frac{m_{\eta^0}^2}{m_i^2}\right)
\right]\equiv \left(y^T\Lambda y\right)_{\alpha\beta},
\end{equation}
where $m_{\eta^0}^2=\left(m_H^2+m_A^2\right)/2$ and $\Lambda$ is the $2\times2$ matrix.

The Yukawa matrix $y$ is parametrized as follows by the Casas-Ibarra parametrization~\cite{Casas:2001sr} in order to accommodate the neutrino oscillation data
\begin{equation}
 y=\sqrt{\Lambda}^{-1}C\sqrt{\hat{m}_{\nu}}U_\mathrm{PMNS}^{\dag},
\label{eq:casas-ibarra}
\end{equation}
where the $2\times3$ matrix $C$ is a complex orthogonal matrix satisfying $CC^T=\1_\mathrm{2\times2}$, which is parametrized as
\begin{align}
 C=&\left(
\begin{array}{ccc}
 0 & \cos\chi & -\sin\chi\\
 0 & \kappa\sin\chi & \kappa\cos\chi
\end{array}
\right)\quad\text{for normal hierarchy},\\
 C=&\left(
\begin{array}{ccc}
 \cos\chi & -\sin\chi & 0\\
 \kappa\sin\chi & \kappa\cos\chi & 0
\end{array}
\right)\quad\text{for inverted hierarchy},
\end{align}
with $\kappa=\pm1$ and a complex angle $\chi$. 
The diagonalized neutrino mass matrix $\hat{m}_{\nu}$ is given by $\hat{m}_{\nu}=\mathrm{diag}\left(\hat{m}_1,\hat{m}_2,\hat{m}_3\right)$, and the PMNS matrix $U_\mathrm{PMNS}$ is 
parametrized as\footnote{Only one Majorana phase exists since one of the neutrino mass eigenvalues remains massless in the models.}
\begin{align}
U_\mathrm{PMNS}=&\left(
\begin{array}{ccc}
 1 & 0 & 0\\
0 & \cos\theta_{23} & \sin\theta_{23}\\
0 & -\sin\theta_{23} & \cos\theta_{23}
\end{array}
\right)\left(
\begin{array}{ccc}
 \cos\theta_{13} & 0 & \sin\theta_{13}e^{-i\delta_\mathrm{CP}}\\
0 & 1 & 0\\
-\sin\theta_{13}e^{i\delta_\mathrm{CP}} & 0 & \cos\theta_{13}
\end{array}
\right)\nonumber\\
&\times\left(
\begin{array}{ccc}
 \cos\theta_{12} & \sin\theta_{12} & 0\\
-\sin\theta_{12} & \cos\theta_{12} & 0\\
0 & 0 & 1
\end{array}
\right)\left(
\begin{array}{ccc}
 1 & 0 & 0\\
0 & e^{i\varphi_\mathrm{CP}} & 0\\
0 & 0 & 1
\end{array}
\right),
\end{align}
where $\theta_{ij}$ is the mixing angles determined by the neutrino oscillation data, $\delta_\mathrm{CP}$ and $\varphi_\mathrm{CP}$ correspond to the Dirac and Majorana phases, respectively.
Since the rank of the Yukawa matrix $y$ is two in the models considered here, one finds that one of the mass eigenvalues of neutrinos is zero, which is the lightest one. 
From the definition of normal and inverted neutrino mass hierarchy, 
the lightest neutrino mass eigenvalue corresponds to $\hat{m}_1=0$ for normal hierarchy and $\hat{m}_3=0$ for inverted hierarchy, respectively. 
The imaginary part of $\sin\chi$, $\delta_{\mathrm{CP}}$, and $\varphi_{\mathrm{CP}}$ are three CP-violating parameters in the models.

The neutrino masses, mixing angles and the Dirac phase are limited in the following range at $3\sigma$ confidence level of the global fit analysis~\cite{Esteban:2018azc},
\begin{align}
&0.275\leq\sin^2\theta_{12}\leq0.350,\qquad
0.418\leq\sin^2\theta_{23}\leq0.627,\qquad
0.02045\leq\sin^2\theta_{13}\leq0.02439,\nonumber\\
&6.79\leq \frac{\Delta m_{21}^2}{10^{-5}~\mathrm{eV}^2} \leq8.01,\qquad
2.427\leq \frac{\Delta m_{31}^2}{10^{-3}~\mathrm{eV}^2}\leq2.625,\qquad
125^\circ\leq \delta_\mathrm{CP} \leq 392^\circ,
\label{eq:nh}
\end{align}
for normal mass hierarchy and 
\begin{align}
&0.275\leq\sin^2\theta_{12}\leq0.350,\qquad
0.423\leq\sin^2\theta_{23}\leq0.629,\qquad
0.02068\leq\sin^2\theta_{13}\leq0.02463,\nonumber\\
&6.79\leq \frac{\Delta m_{21}^2}{10^{-5}~\mathrm{eV}^2} \leq8.01,\quad
-2.611\leq \frac{\Delta m_{32}^2}{10^{-3}~\mathrm{eV}^2}\leq-2.412,\qquad
196^\circ\leq \delta_\mathrm{CP} \leq 360^\circ,
\label{eq:ih}
\end{align}
for inverted mass hierarchy where the squared neutrino mass difference is defined by $\Delta m_{ij}^2\equiv \hat{m}_i^2-\hat{m}_j^2$.

\section{Charged lepton flavor violation}
\label{sec:3}

\begin{figure}[t]
\begin{center}
\includegraphics[scale=1.1]{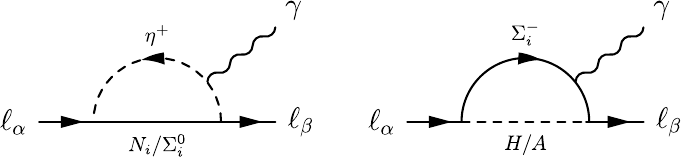}
\caption{Feynman diagrams for the charged lepton flavor violation process $\ell_\alpha\to \ell_\beta\gamma$. The diagrams that a photon is attached to the external legs do not contribute to the amplitude and are omitted.}
\label{fig:lfv}
\end{center}
\end{figure}

The charged LFV processes such as $\ell_\alpha\to\ell_\beta\gamma$, $\ell_\alpha\to3\ell_\beta$, and $\mu$-$e$ conversion give severe constraints on the models. 
In particular, the process $\ell_\alpha\to\ell_\beta\gamma$ is the strongest and is studied in various models. 
In the radiative neutrino mass models we consider, the diagrams relevant to the process $\ell_\alpha\to\ell_\beta\gamma$ are shown in Fig.~\ref{fig:lfv}, and the branching ratio is calculated as 
\begin{align}
 \mathrm{Br}\left(\ell_\alpha\to\ell_\beta\gamma\right)
=&~\frac{3\alpha_\mathrm{em}}{64\pi G_F^2}\left|
\sum_{i=1}^2\frac{y_{i\beta}^{*}y_{i\alpha}}{m_{\eta^+}^2}F_2\bigg(\frac{m_{i}^2}{m_{\eta^+}^2}\bigg)
\right|^2
\mathrm{Br}\left(\ell_\alpha\to\ell_\beta\nu_\alpha\overline{\nu_\beta}\right),
\label{eq:lfv1}
\end{align}
for the SS model~\cite{Toma:2013zsa},
\begin{align}
 \mathrm{Br}\left(\ell_\alpha\to\ell_\beta\gamma\right)
=&~\frac{3\alpha_\mathrm{em}}{64\pi G_F^2}\left|
\frac{y_{2\beta}^{*}y_{2\alpha}}{m_{\eta^+}^2}\left[
F_2\bigg(\frac{m_{2}^2}{m_{\eta^+}^2}\bigg)
-\frac{m_{\eta^+}^2}{m_{2}^2}F_2\left(\frac{m_H^2}{m_{2}^2}\right)
-\frac{m_{\eta^+}^2}{m_{2}^2}F_2\left(\frac{m_A^2}{m_{2}^2}\right)
\right]\right.\nonumber\\
&\hspace{1.7cm}+\left.\frac{y_{1\beta}^{*}y_{1\alpha}}{m_{\eta}^2}F_2\left(\frac{m_{1}^2}{m_{\eta}^2}\right)
\right|^2
\mathrm{Br}\left(\ell_\alpha\to\ell_\beta\nu_\alpha\overline{\nu_\beta}\right),
\label{eq:lfv2}
\end{align}
for the ST model~\cite{Rocha-Moran:2016enp}, and
\begin{align}
 \mathrm{Br}\left(\ell_\alpha\to\ell_\beta\gamma\right)
=&~\frac{3\alpha_\mathrm{em}}{64\pi G_F^2}\left|
\sum_{i=1}^2\frac{y_{i\beta}^{*}y_{i\alpha}}{m_{\eta^+}^2}\left[
F_2\bigg(\frac{m_{i}^2}{m_{\eta^+}^2}\bigg)
-\frac{m_{\eta^+}^2}{m_{i}^2}F_2\left(\frac{m_H^2}{m_{i}^2}\right)
-\frac{m_{\eta^+}^2}{m_{i}^2}F_2\left(\frac{m_A^2}{m_{i}^2}\right)
\right]\right|^2\nonumber\\
&\hspace{1.7cm}\times\mathrm{Br}\left(\ell_\alpha\to\ell_\beta\nu_\alpha\overline{\nu_\beta}\right),
\label{eq:lfv3}
\end{align}
for the TT model~\cite{Chao:2012sz, vonderPahlen:2016cbw} where the loop function $F_2(x)$ is given by
\begin{equation}
F_2\left(x\right)=\frac{1-6x+3x^2+2x^3-6x^2\log{x}}{6(1-x)^4}.
\end{equation}
The branching ratios of the process $\ell_\alpha\to \ell_\beta\nu_\alpha\overline{\nu_\beta}$ appearing in Eqs.~(\ref{eq:lfv1}), (\ref{eq:lfv2}) and (\ref{eq:lfv3}) are determined by experiments~\cite{Zyla:2020zbs}:
\begin{align}
\mathrm{Br}\left(\mu\to e\nu_\mu\overline{\nu_e}\right)\approx&~1,\\
\mathrm{Br}\left(\tau\to\mu\nu_\tau\overline{\nu_\mu}\right)\approx&~0.1739,\\
\mathrm{Br}\left(\tau\to e\nu_\tau\overline{\nu_e}\right)\approx&~0.1782,
\end{align}
while the current upper bounds of the charged LFV processes are summarized as~\cite{Zyla:2020zbs}
\begin{align}
\mathrm{Br}\left(\mu\to e\gamma\right)\leq4.2\times10^{-13},\\
\mathrm{Br}\left(\tau\to e\gamma\right)\leq3.3\times10^{-8},\\
\mathrm{Br}\left(\tau\to \mu\gamma\right)\leq4.4\times10^{-8}.
\end{align}
As can be seen, the constraint on  the charged LFV process $\mu\to e\gamma$ is especially severe.

For all the models, one finds that the branching ratio of the process $\ell_\alpha\to\ell_\beta\gamma$ behaves as 
$\mathrm{Br}\left(\ell_\alpha\to\ell_\beta\gamma\right)\propto G_F^{-2}m_{\eta^+}^{-4}$ when $m_{i}\ll m_{\eta^+,H,A}$. 
On the other hand, when $m_{i}\gg m_{\eta^+,H,A}$, the branching ratio goes as $\mathrm{Br}\left(\ell_\alpha\to\ell_\beta\gamma\right)\propto G_F^{-2}m_{\eta^+}^4m_{i}^{-8}$ for the TT model and $\mathrm{Br}\left(\ell_\alpha\to\ell_\beta\gamma\right)\propto G_F^{-2}m_{i}^{-4}$ for the other models.
This implies that for the TT model, the branching ratio rapidly decreases with the fermion mass $m_i$, and the charged LFV constraints tend to be alleviated compared to the other models.

\section{Charged lepton EDMs at two-loop level}
\label{sec:4}
In the previous work~\cite{Abada:2018zra}, the charged lepton EDMs have been calculated in the minimal scotogenic model which corresponds to the SS model.
We revisit this calculation and extend it to the other models (the ST and TT models). 
In general, the EDM of charged lepton $\ell_\alpha$ can be expressed in the form of 
\begin{align}
d_\alpha/e=&-\frac{m_\alpha}{(4\pi)^4m_{\eta^+}^2}\sum_{i,j,\beta}
\left[
J^{M}_{ij\alpha\beta}\sqrt{\xi_{i}\xi_{j}}
I_{M}\left(\xi_{i},\xi_{j},\xi_H,\xi_A,\xi_\alpha,\xi_\beta,\xi_{\nu_\beta}\right)\right.\nonumber\\
&\hspace{3.1cm}\left.
+J^D_{ij\alpha\beta}I_{D}\left(\xi_{i},\xi_{j},\xi_H,\xi_A,\xi_\alpha,\xi_\beta,\xi_{\nu_{\beta}}\right)
\right],
\label{eq:edm1}
\end{align}
where $m_{\alpha}$ is the mass of $\ell_\alpha$ and $\xi_a\equiv m_a^2/m_{\eta^+}^2$ is the dimensionless parameter defined by the mass ratios. 
The first and second terms in the square bracket of Eq.~(\ref{eq:edm1}) correspond to the Majorana type contribution involving lepton number violation and the Dirac type contribution with lepton number conservation in the loop. Since the masses in the internal fermion propagators are picked up for the Majorana type contribution, it is proportional to $m_im_j$ as can be seen in Eq.~(\ref{eq:edm1}).
The (rephasing-invariant) phase factors $J^{M}_{ij\alpha\beta}$ and $J^D_{ij\alpha\beta}$ are given by
\begin{align}
J^{M}_{ij\alpha\beta}=\mathrm{Im}\left(y_{j\alpha}^{*}y_{j\beta}^{*}y_{i\beta}y_{i\alpha}\right),\qquad
J^{D}_{ij\alpha\beta}=\mathrm{Im}\left(y_{j\alpha}^{*}y_{j\beta}y_{i\beta}^{*}y_{i\alpha}\right),
\end{align}
for Majorana and Dirac type contributions, respectively.
From the above expressions, one can find that the phase factors $J_{ij\alpha\beta}^{M}$ and $J_{ij\alpha\beta}^D$ are anti-symmetric under the exchange $i\leftrightarrow j$. 
This implies that only the anti-symmetric part of the loop functions in Eq.~(\ref{eq:edm1}) contributes to the EDMs if both diagrams exist. 

\begin{figure}[t]
 \begin{center} 
  \includegraphics[scale=0.75]{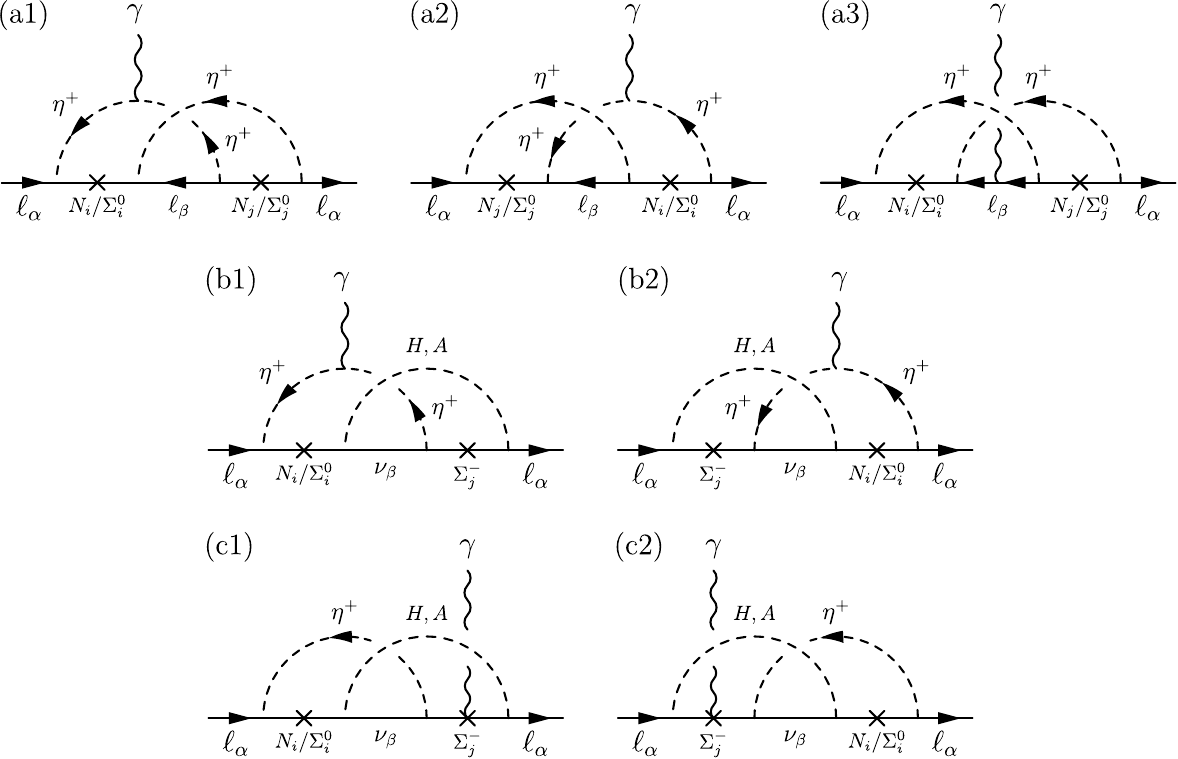}
\caption{Majorana type diagrams contributing to charged lepton EDMs. The cross mark represents the lepton number violation picking up the Majorana mass term of the propagators.}
\label{fig:edm_majorana}
 \end{center}
\end{figure}

\begin{figure}[t]
 \begin{center} 
  \includegraphics[scale=0.75]{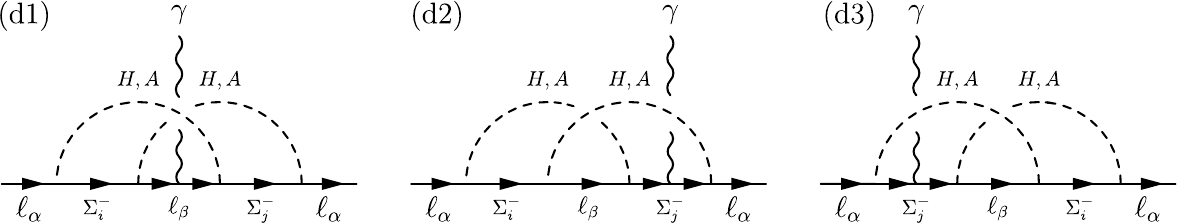}
\caption{Dirac type diagrams contributing to charged lepton EDMs.}
\label{fig:edm_dirac}
 \end{center}
\end{figure}

\begin{figure}[t]
 \begin{center} 
  \includegraphics[scale=0.75]{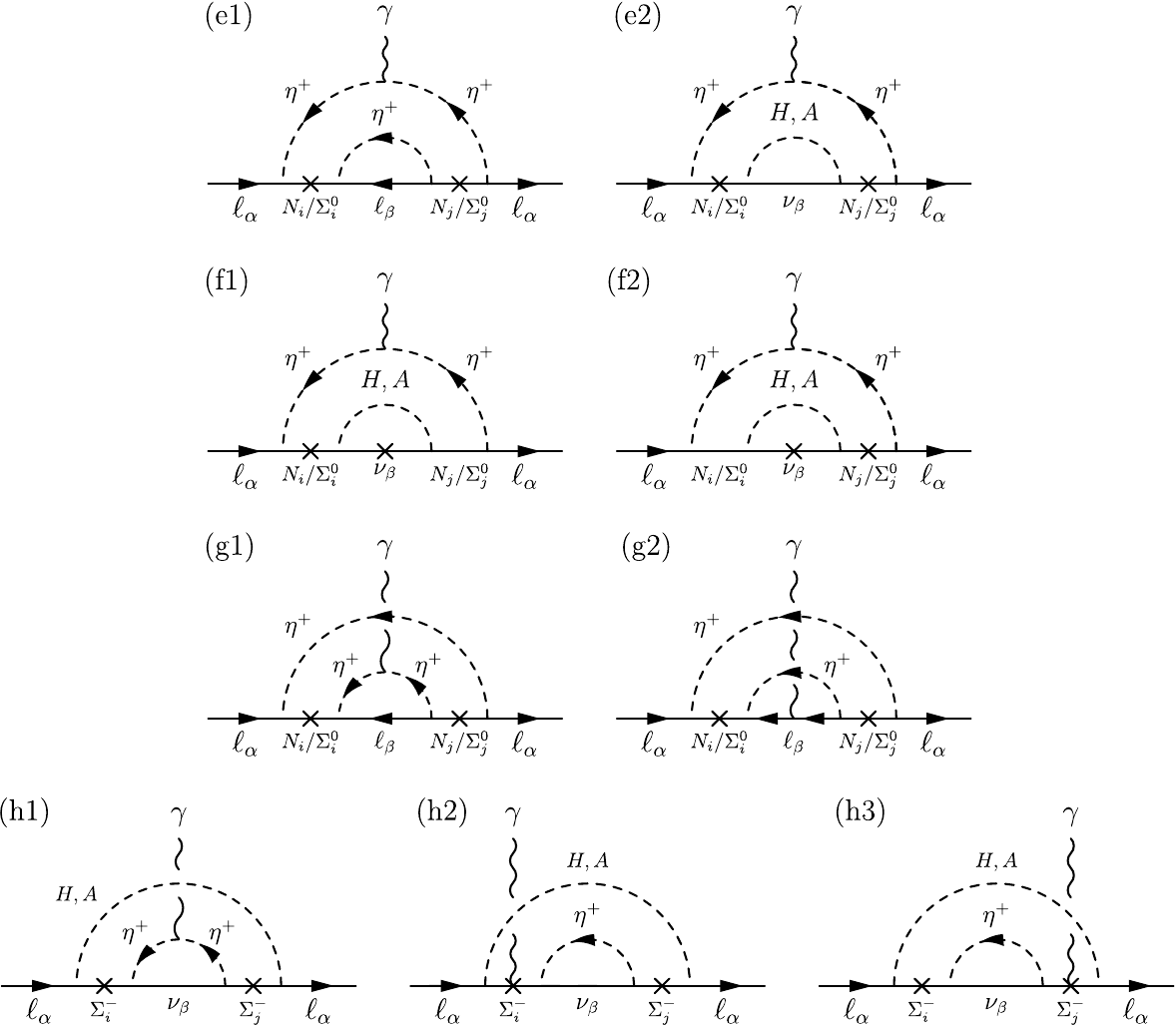}
\caption{Majorana type diagrams not contributing to charged lepton EDMs.}
\label{fig:edm_majorana_zero}
 \end{center}
\end{figure}

\begin{figure}[t]
 \begin{center} 
  \includegraphics[scale=0.75]{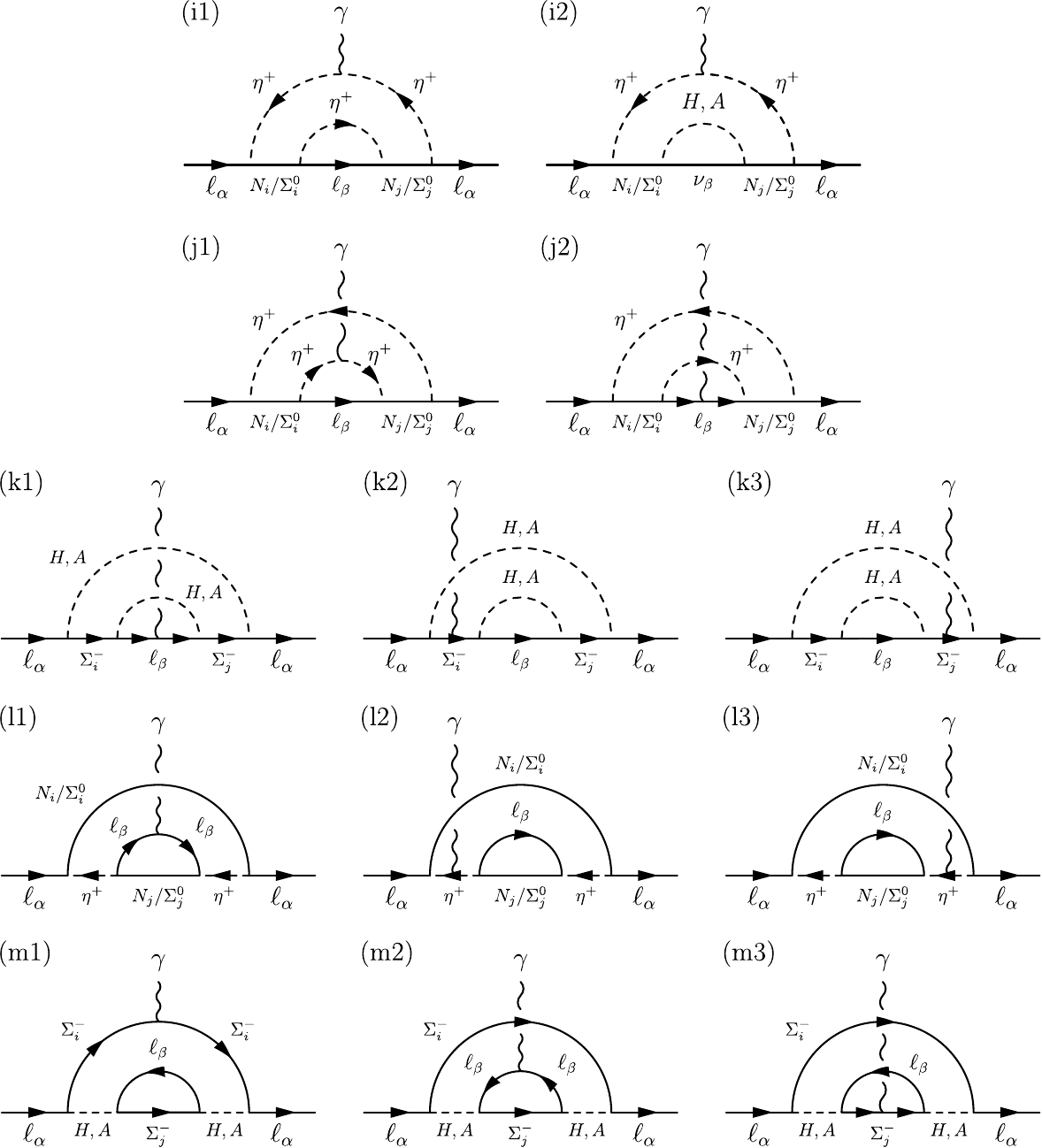}
\caption{Dirac type diagrams not contributing to charged lepton EDMs.}
\label{fig:edm_dirac_zero}
 \end{center}
\end{figure}

The loop functions $I_M$ and $I_D$ in Eq.~(\ref{eq:edm1}) can be calculated from the corresponding diagrams shown in 
Figs.~\ref{fig:edm_majorana} and \ref{fig:edm_dirac}. 
The formulae of the Majorana type diagrams are summarized below: 
\begin{align}
 I_{M}^{\mathrm{(a)}}\left(\xi_i,\xi_j,\xi_\alpha,\xi_\beta\right)=&\int_0^1dxdydz~\delta(x+y+z-1)\int_0^1dsdtdu~\delta(s+t+u-1)\nonumber\\
   &\times \left[\frac{xyt(1-u)(-s+ys-zu)}{D^2\left(\xi_{i},\xi_{j},1,1,\xi_\alpha,\xi_\beta\right)}
+\frac{1}{2}\frac{yu(-xs+zt)(xt-zs)}{D^2\left(\xi_{i},\xi_{j},1,1,\xi_\alpha,\xi_\beta\right)}\right],
\end{align}
\begin{align}
I_{M}^{\mathrm{(b)}}\left(\xi_i,\xi_j,\xi_H,\xi_A,\xi_\alpha,\xi_{\nu_{\beta}}\right)=&\int_0^1dxdydz~\delta(x+y+z-1)\int_0^1dsdtdu~\delta(s+t+u-1)\nonumber\\
&\times\sum_{\varphi=H,A}\frac{xyt(1-u)(-s+ys-zu)}{D^2\left(\xi_{i},\xi_{j},1,\xi_\varphi,\xi_\alpha,\xi_{\nu_\beta}\right)},\\
I_{M}^{\mathrm{(c)}}\left(\xi_i,\xi_j,\xi_H,\xi_A,\xi_\alpha,\xi_{\nu_\beta}\right)=&\int_0^1dxdydz~\delta(x+y+z-1)\int_0^1dsdtdu~\delta(s+t+u-1)\nonumber\\
&\times \sum_{\varphi=H,A}ys\frac{\left((1-y)t+xu\right)(x+zu)-x(1-y)}{D^2\left(\xi_{j},\xi_{i},\xi_\varphi,1,\xi_\alpha,\xi_{\nu_\beta}\right)},
\end{align}
where the superscripts (a), (b), and (c) represent the corresponding Feynman diagrams in Fig.~\ref{fig:edm_majorana}, and the denominator in the above formulae is defined by
\begin{align}
D\left(\xi_{i},\xi_{j},\xi_\varphi,\xi_{\tilde{\varphi}},\xi_\alpha,\xi_\beta\right)\equiv&~
y(1-y)\left(s\xi_{i}+t\xi_\varphi\right)+xu\xi_{j}+zu\xi_{\tilde{\varphi}}+yu\xi_\beta\nonumber\\
&+y(1-y)\left[-\frac{x(1-x)}{y(1-y)}u-t+\left\{t+\frac{xu}{1-y}\right\}^2\right]\xi_\alpha.
\end{align}
For the Dirac type diagrams (d1)-(d3) in Fig.~\ref{fig:edm_dirac}, this contribution can be ignored since it is proportional to the mass splitting between the neutral inert scalars $H$ and $A$, which must be sufficiently small to reproduce the tiny neutrino masses.

Here we have shown the exact formulae of charged lepton EDMs at the two-loop level without neglecting the lepton masses for completeness.\footnote{In a precise sense, the EDM calculation must be done in the neutrino mass eigenbasis and the dependence of the PMNS matrix comes in the EDM formulae. However the calculation in the neutrino flavor basis which has been done in the above has no substantial difference from that in the neutrino mass eigenbasis since the small neutrino masses can eventually be ignored.}
In a realistic study, the charged lepton and neutrino masses can be neglected ($\xi_\alpha,\xi_\beta,\xi_{\nu_\beta}\ll1$), and thus 
the EDM formulae can be further simplified as will be seen below. 

In addition to the above diagrams, the diagrams shown in Figs.~\ref{fig:edm_majorana_zero} and \ref{fig:edm_dirac_zero} might contribute to charged lepton EDMs.
As we will explain shortly, however, we conclude these diagrams induce no charged lepton EDM for some reasons.
The Majorana type diagrams (e1), (e2), (f1), (f2) in Fig.~\ref{fig:edm_majorana_zero} and the Dirac type diagrams (i1) and (i2) in Fig.~\ref{fig:edm_dirac_zero} do not contribute to charged lepton EDMs 
since the resultant loop function is completely symmetric under the exchange $i\leftrightarrow j$. 
For the Majorana type diagrams (g1) and (g2) in Fig.~\ref{fig:edm_majorana_zero}, a non-zero contribution to the charged lepton EDMs arises from each diagram. 
However, we have checked that the sum of these contributions exactly cancels. 
We have checked that the Dirac type diagrams (j1) and (j2) in Fig.~\ref{fig:edm_dirac_zero} also similarly cancel. 
For the Majorana type diagrams (h1)-(h3) and the Dirac type diagrams (k1)-(k3),
we have applied the same arguments with the Ref.~\cite{Shabalin:1978rs} where the quark EDMs in the SM identically vanish at the two-loop level 
using the Lagrangian renormalized at one-loop level,\footnote{Similar argument of the cancellations has been discussed in the Type-I seesaw models~\cite{Ng:1995cs, Archambault:2004td, Chang:2004pba}.
} and we have explicitly shown that the EDM contributions from these diagrams exactly cancel.
The other Dirac type diagrams (l1)-(l3) and (m1)-(m3) also do not contribute to the EDMs since the amplitude is proportional to $|y_{i\alpha}|^2|y_{j\beta}|^2$ and have no imaginary part.

In particular, we are interested in the electron EDM since it is precisely measured compared to the other charged leptons. 
Ignoring the charged lepton and neutrino masses in the loop functions and assuming that the CP-even and CP-odd neutral scalar masses are degenerate with the charged scalar mass,\footnote{The degeneracy of the scalar masses is actually required to obtain the small neutrino masses consistently with the neutrino oscillation data.} 
the formula of the electron EDM is simplified as
\begin{align}
 d_e/e\approx&-\frac{m_eJ_M}{(4\pi)^4m_{\eta^+}^2}
\sqrt{\xi_1\xi_2}\tilde{I}_M\left(\xi_1,\xi_2\right),
\end{align} 
where the Majorana type phase factor for an electron is given by $J_M=\sum_{\beta}J_{12e\beta}^{M}$, 
and the modified Majorana type loop function $\tilde{I}_M(\xi_1,\xi_2)$ is explicitly given by
\begin{align}
\tilde{I}_M\left(\xi_1,\xi_2\right)=&~I_M^{\mathrm{(a)}}\left(\xi_1,\xi_2,0,0\right)-I_M^{\mathrm{(a)}}\left(\xi_2,\xi_1,0,0\right),\label{eq:loop_ss}
\end{align}
for the SS model, 
\begin{align}
\tilde{I}_M\left(\xi_1,\xi_2\right)=&~I_M^{\mathrm{(a)}}\left(\xi_1,\xi_2,0,0\right)-I_M^{\mathrm{(a)}}\left(\xi_2,\xi_1,0,0\right)\nonumber\\
&+I_M^{\mathrm{(b)}}\left(\xi_1,\xi_2,1,1,0,0\right)+I_M^{\mathrm{(c)}}\left(\xi_1,\xi_2,1,1,0,0\right),
\label{eq:loop_st}
\end{align}
for the ST model, and 
\begin{align}
\tilde{I}_M\left(\xi_1,\xi_2\right)=&~I_M^{\mathrm{(a)}}\left(\xi_1,\xi_2,0,0\right)-I_M^{\mathrm{(a)}}\left(\xi_2,\xi_1,0,0\right)\nonumber\\
&+I_M^{\mathrm{(b)}}\left(\xi_1,\xi_2,1,1,0,0\right)-I_M^{\mathrm{(b)}}\left(\xi_2,\xi_1,1,1,0,0\right)\nonumber\\
&+I_M^{\mathrm{(c)}}\left(\xi_1,\xi_2,1,1,0,0\right)-I_M^{\mathrm{(c)}}\left(\xi_2,\xi_1,1,1,0,0\right),\label{eq:loop_tt}
\end{align}
for the TT model. For the ST model, we emphasize that there are no $\xi_1\leftrightarrow\xi_2$ exchanged contributions from the (b) and (c) diagrams since the interactions for the singlet and triplet fermions are asymmetric. 
As a result, the predicted electron EDM in the ST model can be larger than the other models as we will discuss in the next section.

Although we have introduced only two fermions in the models, one can consider extensions with more fermions. 
In these cases, the EDM formulae we have calculated can straightforwardly be derived.

\section{Numerical calculation}
\label{sec:5}
\subsection{The loop functions}

\begin{figure}[t]
\begin{center}
 \includegraphics[scale=0.65]{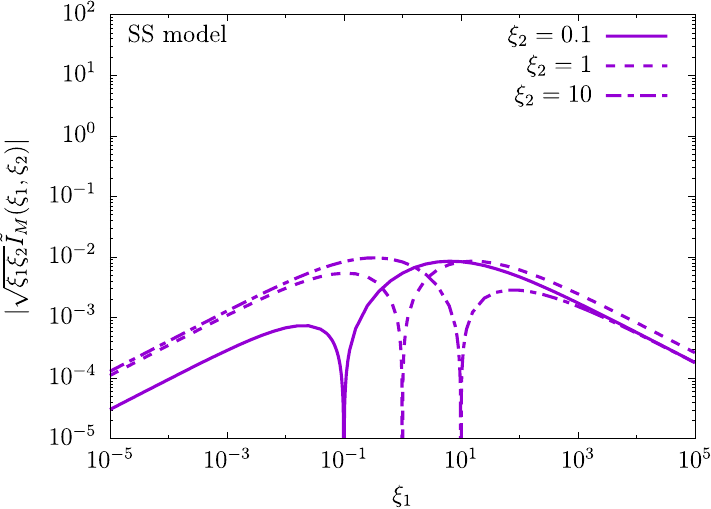}
 \includegraphics[scale=0.65]{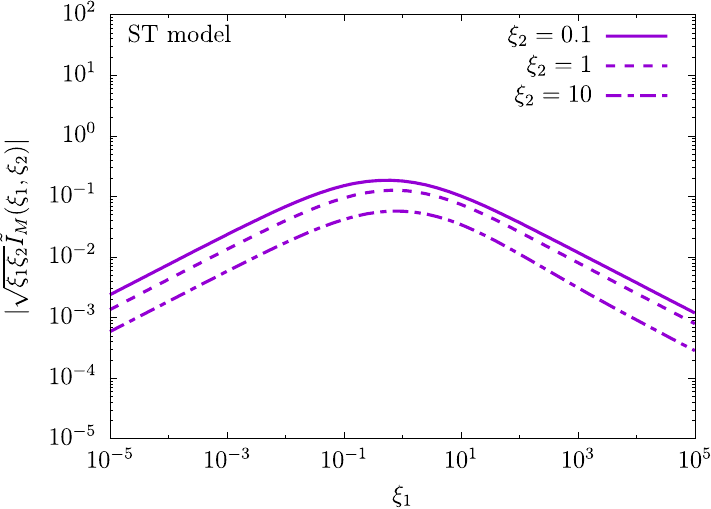}\\
 \includegraphics[scale=0.65]{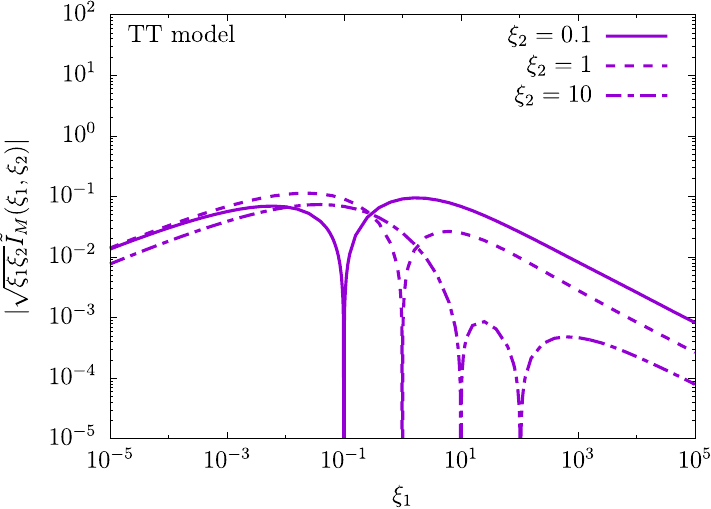}
\caption{Numerical evaluation of the loop functions where $\xi_2$ is fixed as $\xi_2=0.1,1,10$ in each plot and the other parameters are fixed as $\xi_H=\xi_A=1$ and $\xi_\alpha=\xi_\beta=\xi_{\nu_\beta}=0$.}
\label{fig:loop_f}
\end{center}
\end{figure}

The numerical evaluation of the loop functions is shown in Fig.~\ref{fig:loop_f} where the scalar masses are assumed to be degenerate ($\xi_H=\xi_A=1$) and 
the lepton masses are ignored ($\xi_\alpha=\xi_\beta=\xi_{\nu_\beta}=0$).
Thus the remaining parameters in the loop functions are only $\xi_{1}$ and $\xi_{2}$ in this case. 
In Fig.~\ref{fig:loop_f}, the parameter $\xi_2$ is fixed to be $\xi_2=0.1,1,10$.
In the SS and TT models, the loop functions vanish at $\xi_1=\xi_2$. This is because the loop functions shown in Eqs.~(\ref{eq:loop_ss}) and (\ref{eq:loop_tt}) are anti-symmetric under the exchange $i\leftrightarrow j$. On the other hand, the loop function for the ST model given in Eq.~(\ref{eq:loop_st}) is not anti-symmetric and has no zero point of the loop function.
There is another blind spot around $\xi_1\sim100$ in the bottom panel in Fig.~\ref{fig:loop_f}. 
This is due to an accidental cancellation between the loop functions of the diagrams (a), (b), and (c). 

In the numerical calculations, we have found that the (c) diagrams give a dominant contribution. 
Furthermore, we have numerically found that the loop functions behave as $\tilde{I}_M(\xi_1,\xi_2)\propto \xi_1^{-1}$ when $\xi_1\gg1$, which can be seen in Fig.~\ref{fig:loop_f}. 
One can see that the loop function in the ST model tends to be larger than that in the other models when $\xi_1\gtrsim 0.1$. 

\subsection{The electron EDM}
\begin{figure}[t]
\begin{center}
 \includegraphics[scale=0.59]{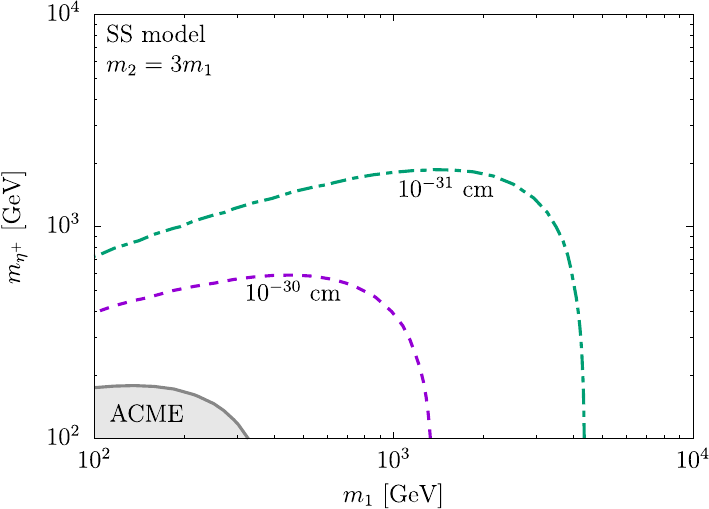}
 \includegraphics[scale=0.59]{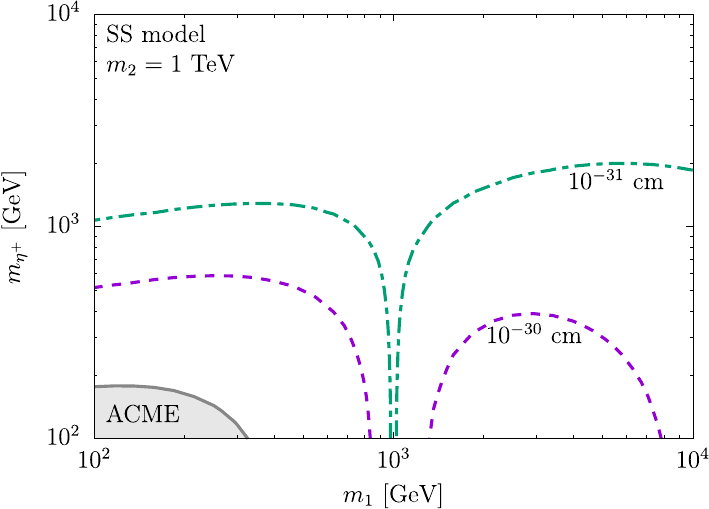}\\
 \includegraphics[scale=0.59]{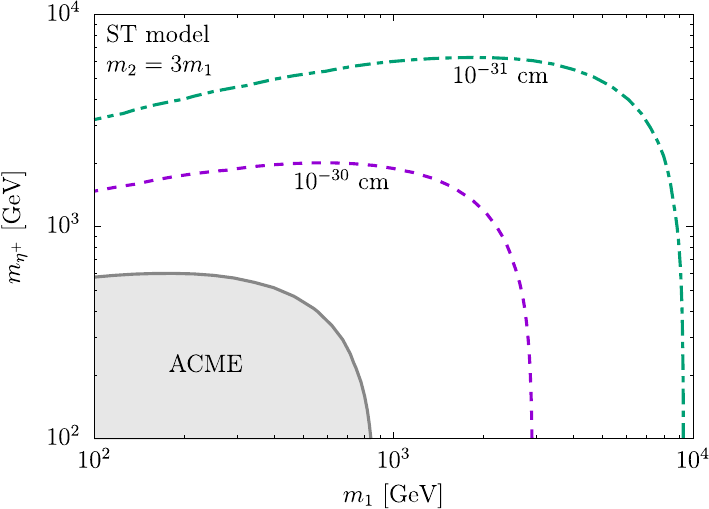}
 \includegraphics[scale=0.59]{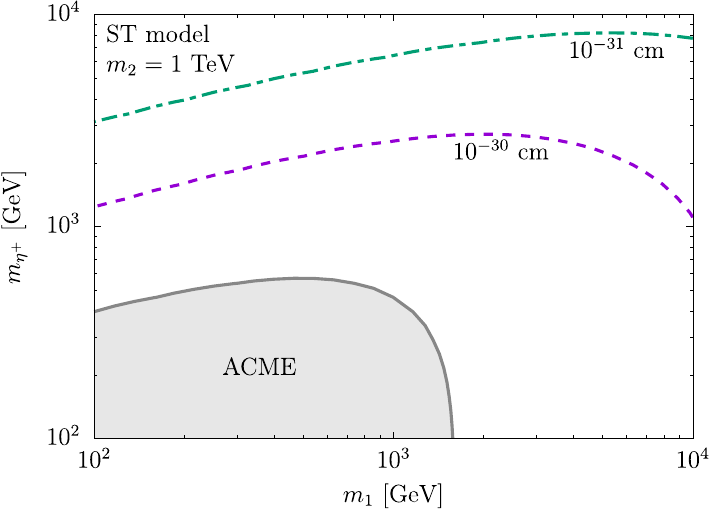}\\
 \includegraphics[scale=0.59]{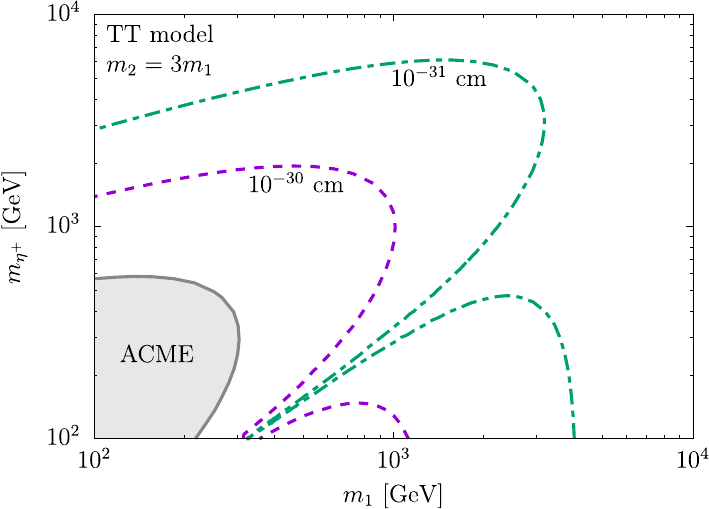}
 \includegraphics[scale=0.59]{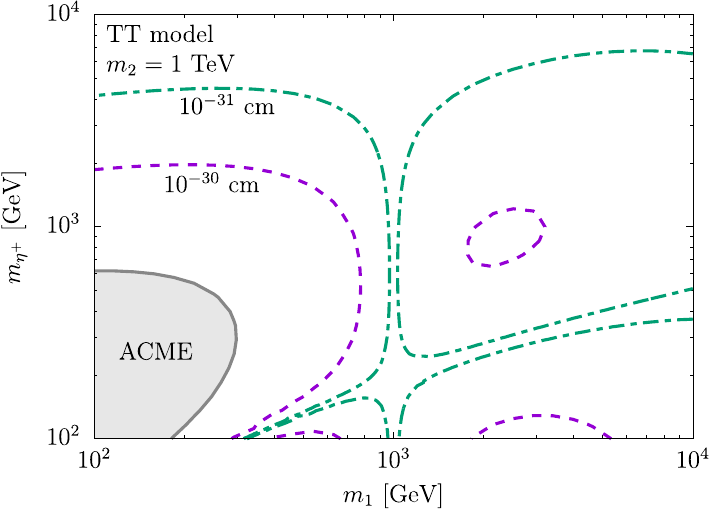}
\caption{Contours of the predicted electron EDM in the ($m_1$, $m_{\eta^+}$) plane where the phase factors $J_M=0.1$ is assumed. The other fermion mass is fixed as $m_2=3m_1$ in the left panels and $m_2=1~\mathrm{TeV}$ in the right panels. The gray region is excluded by the current upper bound of electron EDM given by the ACME Collaboration~\cite{Andreev:2018ayy}.}
\label{fig:edm_cont_m}
\end{center}
\end{figure}

The current upper bound of the electron EDM is given by the ACME Collaboration~\cite{Andreev:2018ayy}:
\begin{equation}
 |d_e|/e<1.1\times10^{-29}~\text{cm},
\end{equation}
at 90$\%$ confidence level. 
The future sensitivity is expected to reach up to $|d_e|/e=\mathcal{O}(10^{-30})~\mathrm{cm}$~\cite{Kara:2012ay, edm_future}.
There are also the experimental upper bounds on muon and tau EDMs. However, these bounds are much weaker than the electron case. Thus we focus on the electron EDM.

The contours of the electron EDM are shown in Fig.~\ref{fig:edm_cont_m} where the second fermion mass is fixed to be $m_2=3m_1$ in the left panels and $m_2=1~\mathrm{TeV}$ in the right panels as examples. 
The Majorana type phase factor defined by $J_M\equiv\sum_{\beta}J_{12e\beta}^M$ is also fixed to be $J_M=0.1$ in Fig.~\ref{fig:edm_cont_m}. 
The mass difference between the new scalar particles and the lepton masses are ignored. 
The gray-colored region is excluded by the current upper bound of the ACME Collaboration~\cite{Andreev:2018ayy}. 
The purple dashed and green dot-dashed lines are the contours of $|d_e|/e=10^{-30}$ and $10^{-31}~\mathrm{cm}$.
In the right top and right bottom panels, one can see that the predicted electron EDM vanishes if two fermion masses are degenerate as can be expected from the structure of the loop functions. 
In addition, in the bottom panels, another blind parameter space of electron EDM can be observed. 
These are due to the cancellation between the loop functions as we have seen in the previous subsection. 
In the middle panels (ST model), it can be seen that the predicted electron EDM is larger than the other models. 
This is because there is no negative contribution from the (b) and (c) diagrams which give partial cancellation as can be seen in Eq.~(\ref{eq:loop_st}). 

\begin{figure}[t]
\begin{center}
\includegraphics[scale=0.59]{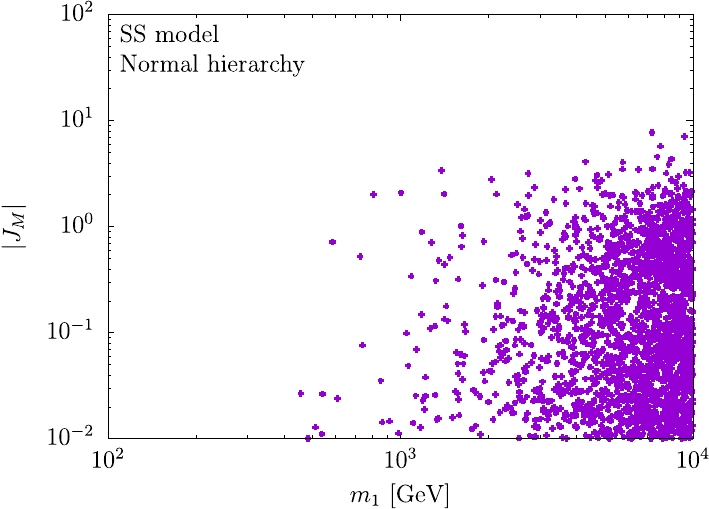}
\includegraphics[scale=0.59]{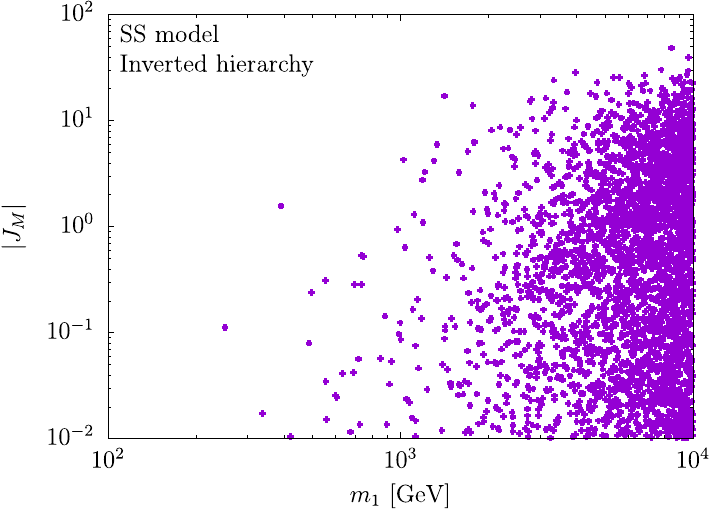}\\
\includegraphics[scale=0.59]{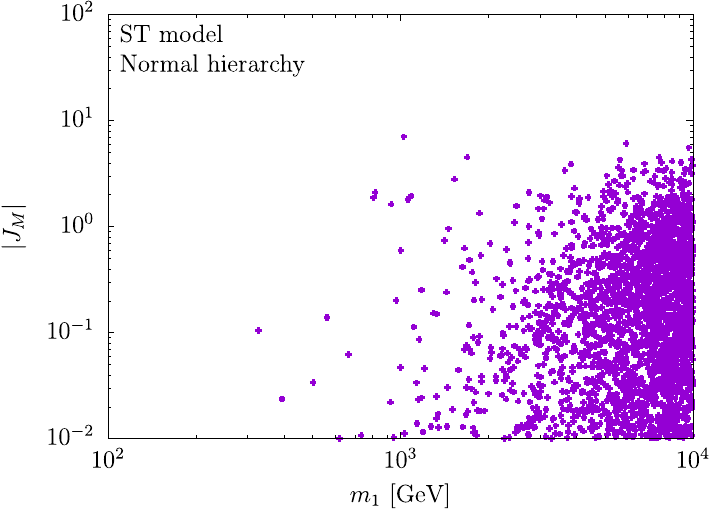}
\includegraphics[scale=0.59]{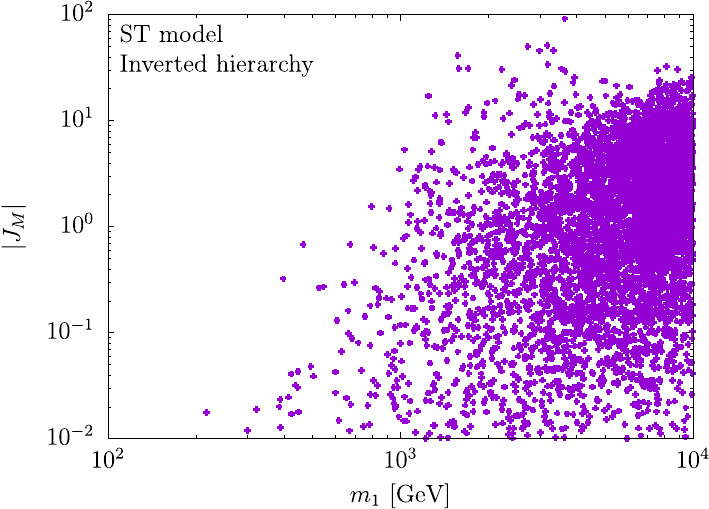}\\
\includegraphics[scale=0.59]{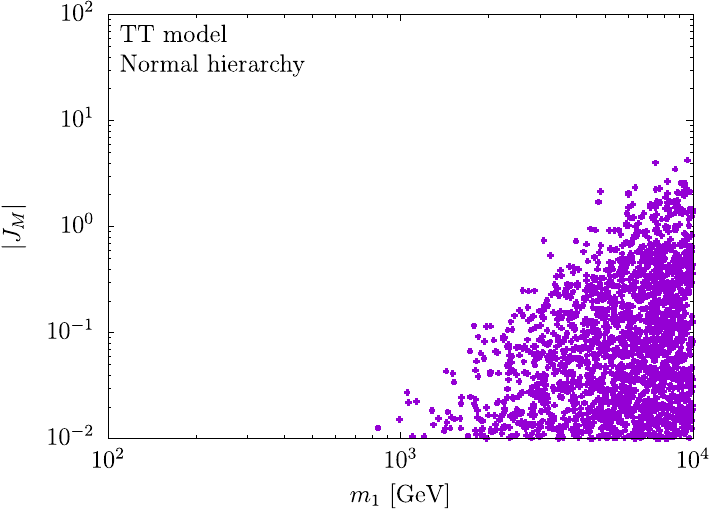}
\includegraphics[scale=0.59]{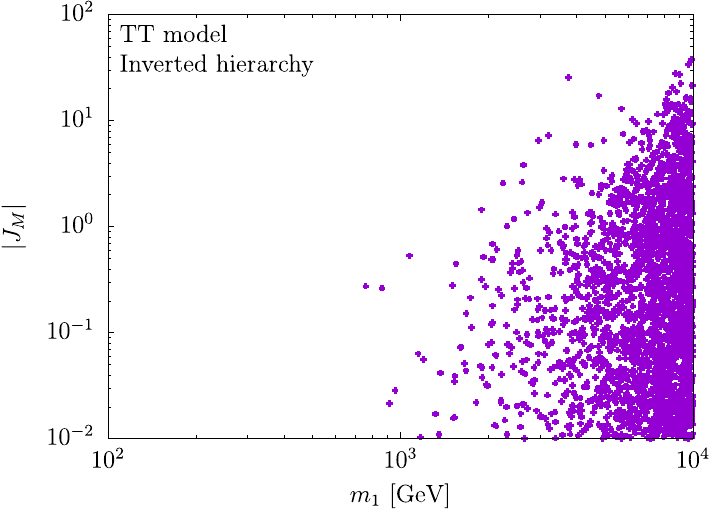}
\caption{Magnitude of the Majorana type phase factor $J_M$ for normal and inverted hierarchy where the constraints of the neutrino oscillation experiments and the charged LFV are taken into account with the scalar coupling $\lambda_5=10^{-10}$.}
\label{fig:phase}
\end{center}
\end{figure}

\begin{figure}[t]
\begin{center}
 \includegraphics[scale=0.60]{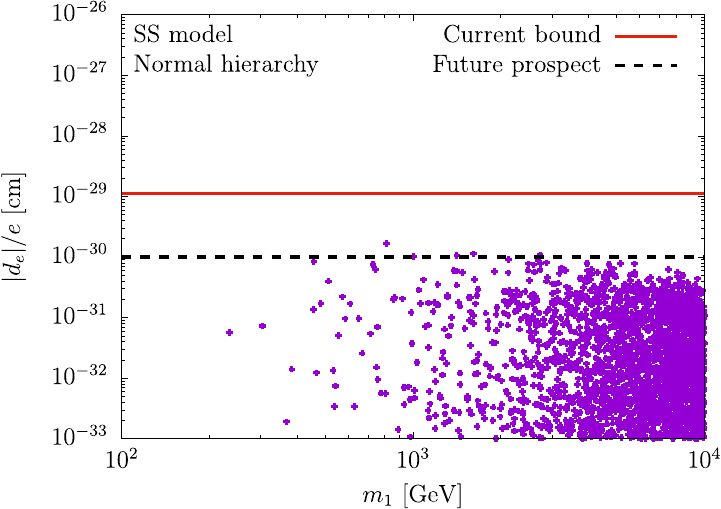}
 \includegraphics[scale=0.60]{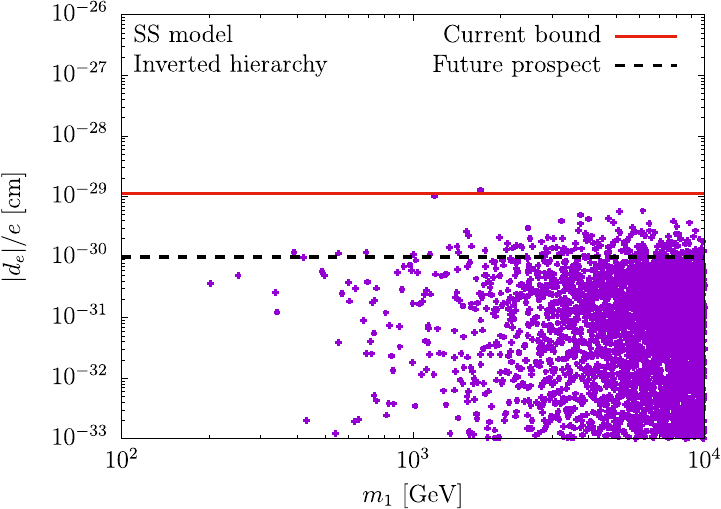}\\
 \includegraphics[scale=0.60]{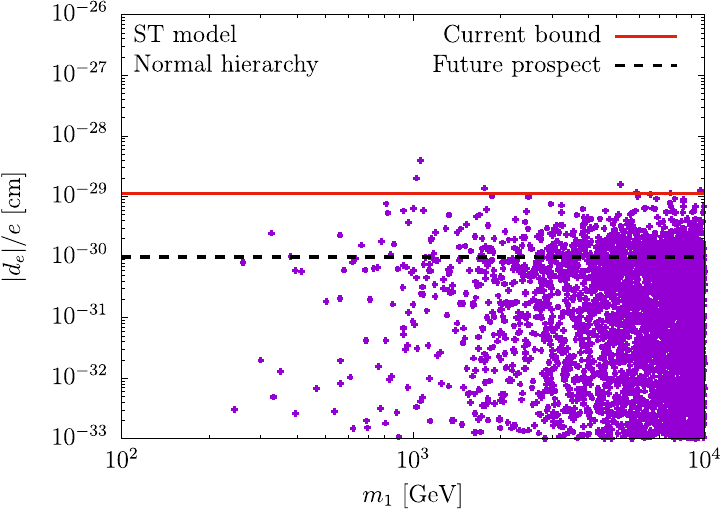}
 \includegraphics[scale=0.60]{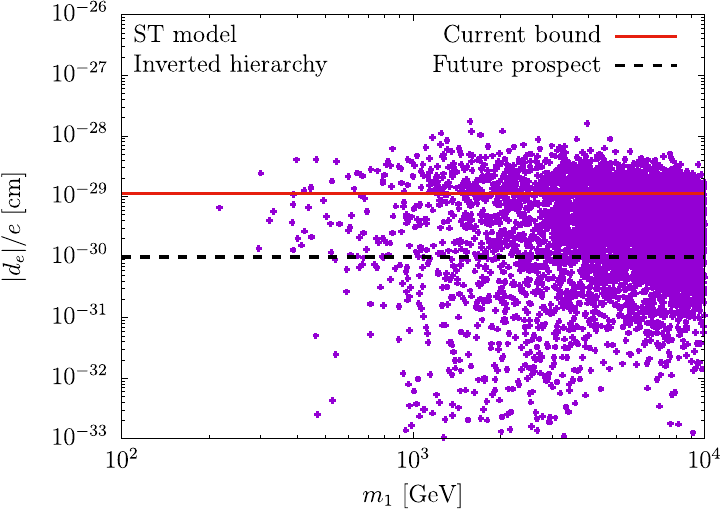}\\
 \includegraphics[scale=0.60]{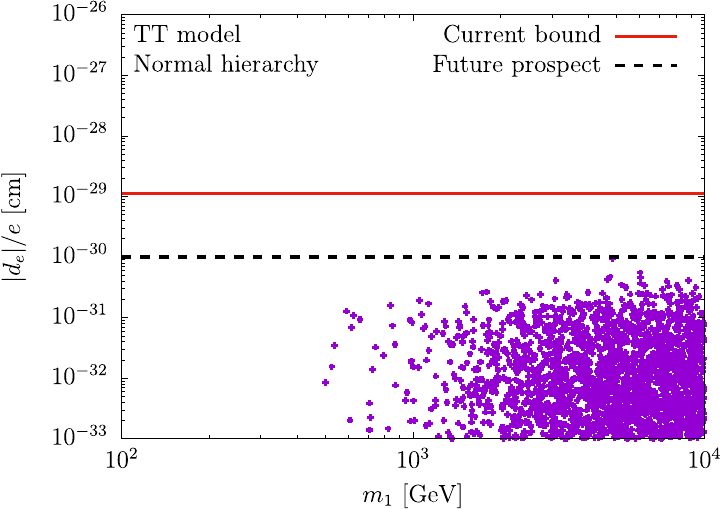}
 \includegraphics[scale=0.60]{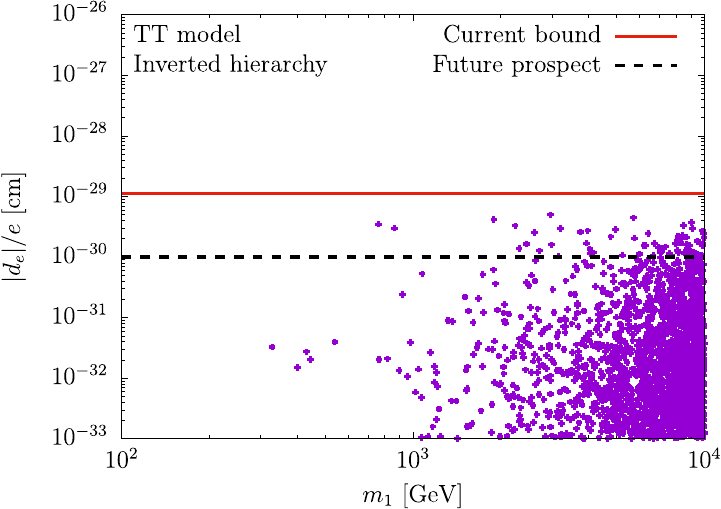}
\caption{Predicted electron EDM with the parameters accommodating neutrino oscillation data and charged LFV constraints. 
The solid red and the dashed black lines represent the current experimental bound of the ACME Collaboration~\cite{Andreev:2018ayy} 
and the future sensitivity~\cite{Kara:2012ay, edm_future}, respectively. The coupling $\lambda_5$ in the scalar potential is fixed to be $\lambda_5=10^{-10}$.}
\label{fig:edm}
\end{center}
\end{figure}

In the following, we give the numerical evaluation of the electron EDM consistent with the charged LFV constraints and the neutrino oscillation experiments using the Casas-Ibarra parametrization given in Eq.~(\ref{eq:casas-ibarra}). 
Since there are many parameters in the models, we randomly scan all the parameters and try to find the maximum magnitude of the electron EDM predicted in the models. 
The intervals of the parameters are taken from the range
\begin{align}
&\hspace{1.5cm}0\leq\varphi_\mathrm{CP}<2\pi,\qquad
0\leq|\sin\chi|\leq1,\nonumber\\
&100~\mathrm{GeV}\leq m_{\eta^+}\leq 10~\mathrm{TeV},\quad
10^{-1} \leq m_{i}/m_{\eta^+} \leq 10.
\end{align}
The neutrino masses and mixing angles are also randomly taken in the $3\sigma$ range in Eq.~(\ref{eq:nh}) for normal hierarchy and Eq.~(\ref{eq:ih}) for inverted hierarchy of the neutrinos.

The absolute value of the Majorana type phase factor $|J_M|$ consistent with the neutrino oscillation experiments and the charged LFV processes is shown in Fig.~\ref{fig:phase} for the case of normal and inverted hierarchy in all the models with the scalar coupling $\lambda_5=10^{-10}$ as a benchmark. 
Note that the Yukawa coupling depends on $y_{i\alpha}\propto\lambda_5^{-1/2}$ via the Casas-Ibarra parametrization in Eq.~(\ref{eq:casas-ibarra}). 
In addition, we restrict the range of the Yukawa coupling $|y_{i\alpha}|<\sqrt{4\pi}$ via the perturbative unitarity. 
From Fig.~\ref{fig:phase}, one finds that the Majorana type phase factor for inverted hierarchy tends to be larger than that for normal hierarchy. 

The predicted electron EDM is shown in Fig.~\ref{fig:edm} for normal and inverted hierarchy in all the models in a similar way to Fig.~\ref{fig:phase}. 
Since the phase factor $J_M$ becomes larger for inverted hierarchy, the predicted electron EDM also tends to be larger than that for normal hierarchy. 
In particular, some parameter sets in the ST model for inverted hierarchy exceeds the current bound of the electron EDM. 
The predicted electron EDM can reach up to the future sensitivity for both normal and inverted hierarchy in the ST model. 
For the other models, some parameter sets can reach up to the future sensitivity for inverted hierarchy while it cannot reach for normal hierarchy.

\section{Conclusions}
\label{sec:6}

Theoretical explanations for small neutrino masses and the existence of dark matter are still unknown though there is overwhelming experimental evidence for them. 
One way for solving these problems simultaneously is the radiative neutrino mass models where the neutrino masses are generated via loop diagrams involving a dark matter candidate with a mass of electroweak scale or TeV scale. 
We have considered the minimal scotogenic model and its extensions with one or two triplet fermions instead of the singlet fermions in the original model. 
These extensions with triplet fermions are motivated from the viewpoint of gauge coupling unification and rich phenomenology. 

In these models, we have calculated the charged lepton EDMs at the two-loop level. 
First, we have categorized all the Feynman diagrams relevant to the charged lepton EDMs. We found that only a few diagrams give a non-zero contribution to the EDMs, and the contributions from the other diagrams exactly cancel. 
We have explicitly checked that this cancellation can be understood from the same arguments of vanishing quark EDMs at two-loop level in the SM. 
In fact, there may be deeper physical meaning of the cancellation. However, this is not main subject of our paper and left for future work~\cite{prep}. 

Second, we have numerically evaluated the analytic formulae of the electron EDM to compare the predictions with the current bound and the future sensitivity given by the ACME Collaboration. 
We studied both normal and inverted hierarchy of neutrinos by taking into account the neutrino oscillation data, the charged LFV constraints, and the perturbative unitarity bound for the Yukawa couplings.
The results show that rather large electron EDM is predicted for inverted hierarchy in all the models we have considered. 
In addition, the genuine new result we have found in the paper is that the ST model where one singlet and one triplet fermions are introduced has predicted an electron EDM larger than the other models. 
This is because the loop function in the ST model is not totally anti-symmetric unlike the other models, which prevents the partial cancellation of the EDM. 
Some parameter space has already been excluded by the current electron EDM bound and a part of the other parameter space can be explored by the future electron EDM experiments. 
Although we have studied the models with only two fermions, one may consider general extensions with arbitrary numbers of singlet fermions and triplet fermions. 
Even in such cases, the EDM formulae can be derived straightforwardly by using our results.

\section*{Acknowledgments}
\noindent
This work was supported by JSPS Grant-in-Aid for Scientific Research KAKENHI Grant No. JP20J12392 (MF), JP20H01895 (JH), and JP20K22349 (TT).
The work of J.H. was supported by Grant-in-Aid for Scientific research from the Ministry of Education, Science, Sports, and Culture (MEXT), Japan (Grant Numbers 16H06492). 
The work of J.H. was also supported by JSPS Core-to-Core Program (Grant Numbers JPJSCCA20200002), 
and World Premier International Research Center Initiative (WPI Initiative), MEXT, Japan.
Numerical computation in this work was carried out at the Yukawa Institute Computer Facility.


\end{document}